\documentclass{cimento}

\setcopyright{CERN on behalf the ALICE Collaboration}
\usepackage{graphicx}  
\usepackage{amsmath}
\usepackage{subcaption}	

\title{Study of charged-particle multiplicities with ALICE}
\author{V.~Zaccolo, on behalf of the ALICE Collaboration}
\instlist{\inst INFN - Turin, Italy}

\newcommand{\s}{$\sqrt{s}$}
\newcommand{\sNN}{$\sqrt{s_{\rm NN}}$}
\newcommand{\dndeta}{d$N_{\rm ch}$/d$\eta$}
\newcommand{\pn}{P$(N_{\rm ch})$}
\newcommand{\avNpart} {$\langle N_\mathrm{part} \rangle$}
\newcommand{\dndetape} {\dndeta/\avNpart}

\begin{document}

\maketitle

\begin{abstract}
The multiplicity measurements include the pseudorapidity density, \dndeta, and the probability distribution as a function of the number of charged particles, \pn. ALICE has measured the multiplicities for three collision systems, for proton-proton, proton-lead and lead-lead collisions at Run 1 and 2 at the Large Hadron Collider. A selection of these results will be presented in these proceedings, concluding with an overview of new measurements planned.
\end{abstract}

\section{Introduction}
\label{Intro}
The multiplicity of charged particles is one of the fundamental observables to describe the global properties of the interactions and are, generally, measured at the start of data taking or later, when the detector is better known or with better analysis techniques.
The available multiplicity results obtained by ALICE will be shown, covering almost all LHC Run 1 and 2 collision energies.
The detector is described elsewhere in detail \cite{Aamodt:2008zz, Abelev:2014ffa}. For the event selection of these results, the forward scintillators V0 are used, as well as the Silicon Pixel Detector (SPD) and the Alice Diffractive detector for Run 2. For the analysis, instead, the Inner Tracking System (ITS) and the Time Projection Chamber (TPC) detectors are used for measurements at midrapidity, while the Forward Multiplicity Detector (FMD) is used to extend the rapidity coverage. 

\section{Results}

\subsection{Proton-proton collisions}
\label{pp}
\begin{figure}[t]
    \begin{subfigure}[h]{0.38\textwidth}
        \hspace{1cm}
        \includegraphics[width=\textwidth]{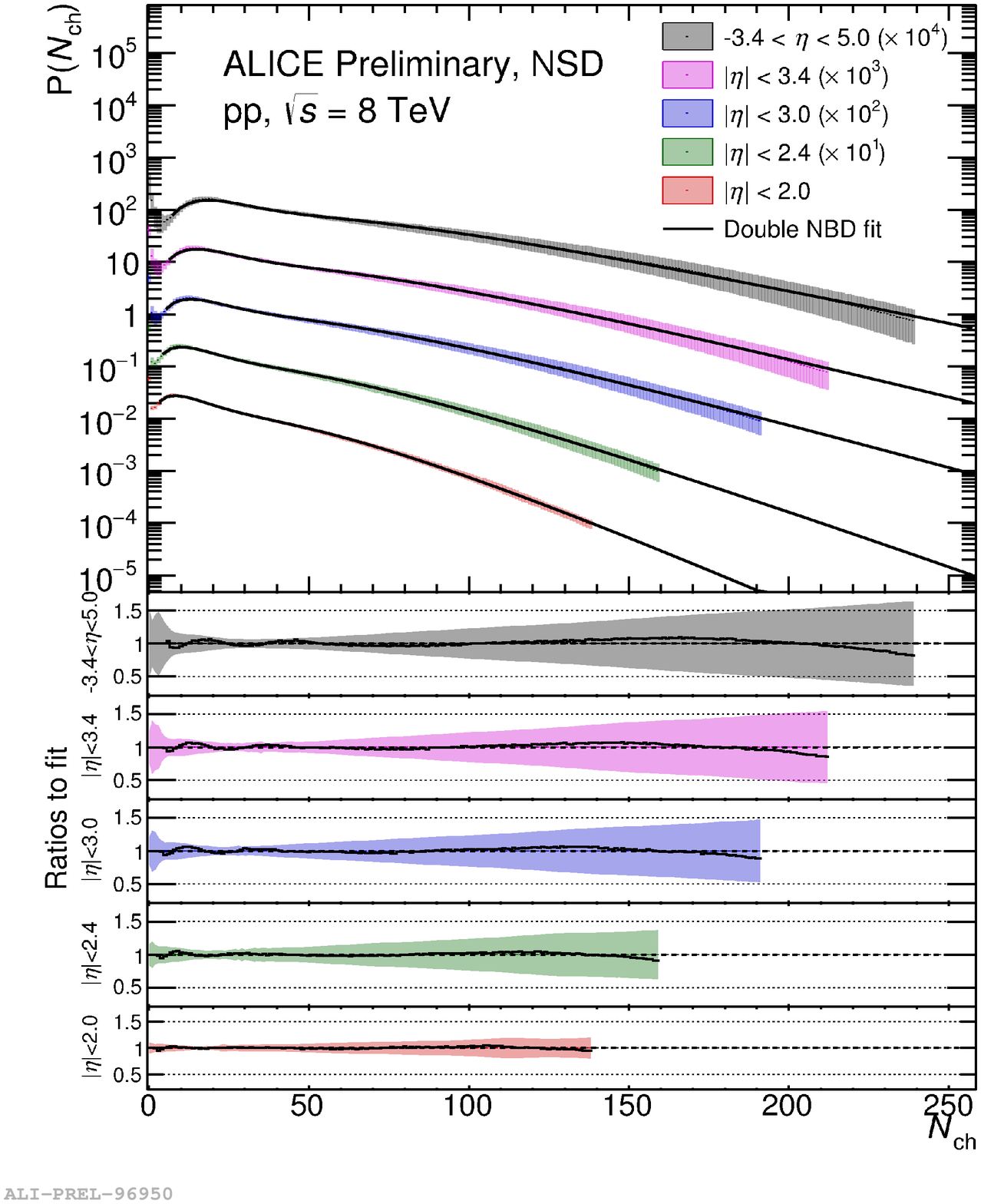}
    \end{subfigure}
  \begin{subfigure}[h]{0.48\textwidth}
       \hspace{1cm}
        \includegraphics[width=\textwidth]{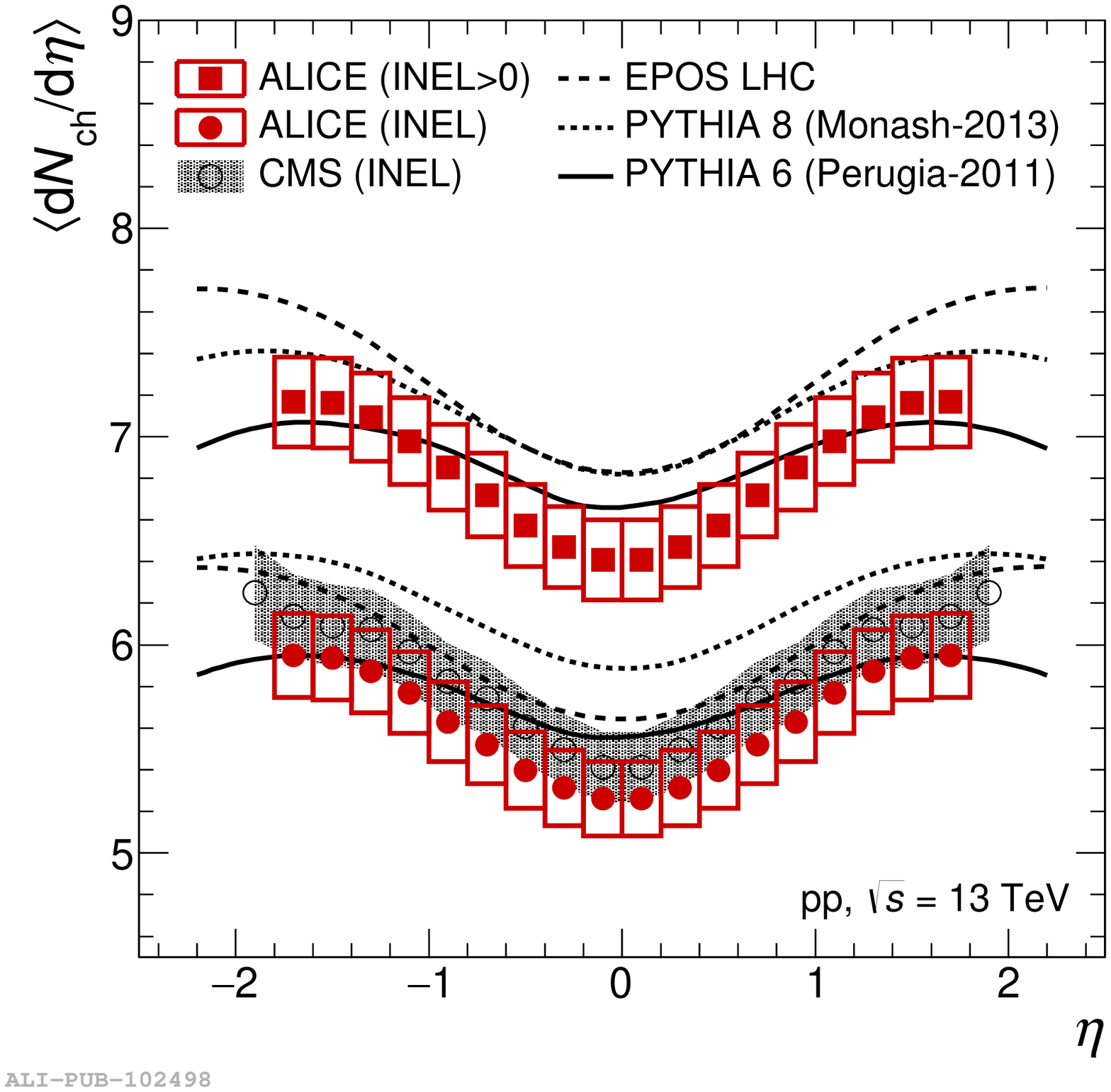}
	\end{subfigure}
\caption{Left: charged-particle multiplicity distributions for NSD pp collisions at \s=8~TeV. The lines show fits to double NBDs \cite{Zaccolo:2015udc}. Right: average pseudorapidity density of charged particles as a function of η produced in pp collisions at s = 13 TeV \cite{Adam:2015pza}.}\label{fig:1}	
\end{figure}

Several sets of measurements have been performed by ALICE at the Run 1 energies for pp collisions. 
Regarding the \pn~distributions as a function of the number of charged particles an in-depth study using both the ITS and the TPC was performed recently \cite{Adam:2015gka} at mid-rapidity. In these measurements, the unfolding technique is used to retrieve the primary charged-particles spectrum. 
At the probed energies, \s = 0.9 to 8 TeV, the usual parameterization with a Negative Binomial Distribution (NBD) does not suffice, instead the sum of two NBDs: $\alpha \text{P}_{\text{NBD}}(n,\langle n\rangle_{\text{1}},k_{\text{1}})+(1-\alpha)\text{P}_{\text{NBD}}(n,\langle n\rangle_{\text{2}},k_{\text{2}})$ fits properly.
This can also be observed in the measurement at wider pseudorapidity ranges, up to $-3.4<\eta<5.0$ in Fig. \ref{fig:1} (left) obtained using the SPD and the FMD \cite{Zaccolo:2015udc}. 
Widening the phase space allows to count nearly twice the number of charged particles. \\ 
Moving to Run 2, results at 13 TeV have been published by ALICE in $|\eta|<1.8$.
In Fig. \ref{fig:1} (right) one can see the \dndeta~distribution for two event classes, inelastic (INEL) and INEL$>0$, where one charged particle is requested in $|\eta|<1$ \cite{Adam:2015pza}. This particular event class allows us to remove a big fraction of Single-Diffractive events and to reduce the uncertainties coming from event trigger inefficiency.

\subsection{Lead-Lead collisions}
\label{PbPb}

ALICE has published multiplicity results both for \mbox{Run 1} Pb--Pb collisions and for Run 2. 
The results for Run 2 at energy in the center of mass \sNN = 5.02 TeV are published as a function of the average number of participants in the collision, \avNpart, calculated using a Monte Carlo Glauber model \cite{Adam:2015ptt}. 
Figure \ref{fig:2} (left) shows the pseudorapidity density normalized to the number of participants, which grows by a factor of about 1.8 going to most central values, where \avNpart~is bigger. The ratio to the results at 2.76 TeV is flat within uncorrelated uncertainties.

\subsection{Proton-Lead collisions}
\label{pPb}

\begin{figure}[t]
    \begin{subfigure}[h]{0.43\textwidth}
        \includegraphics[width=\textwidth]{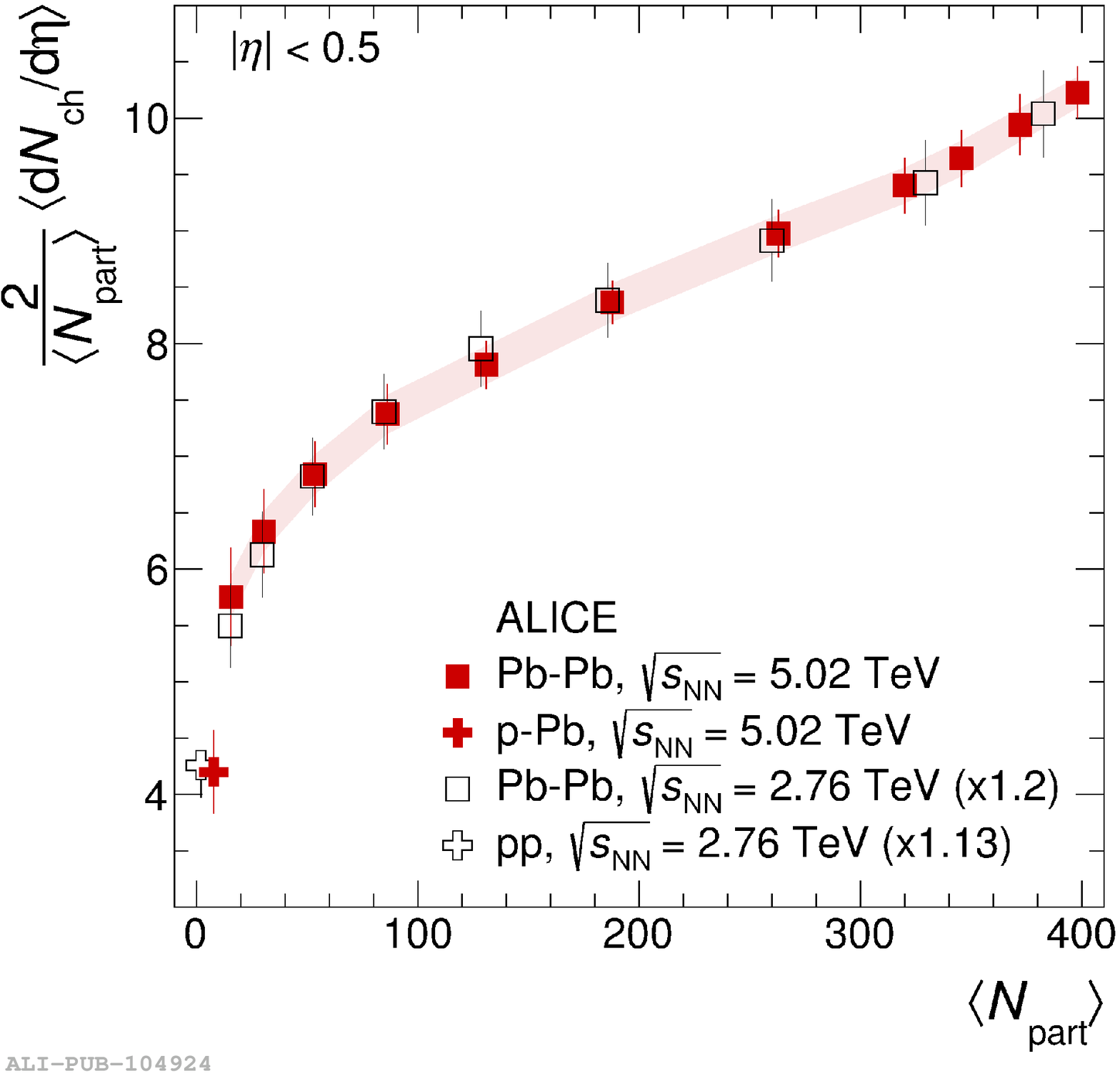}
    \end{subfigure}
  \begin{subfigure}[h]{0.41\textwidth}
       \hspace{1cm}
        \includegraphics[width=\textwidth]{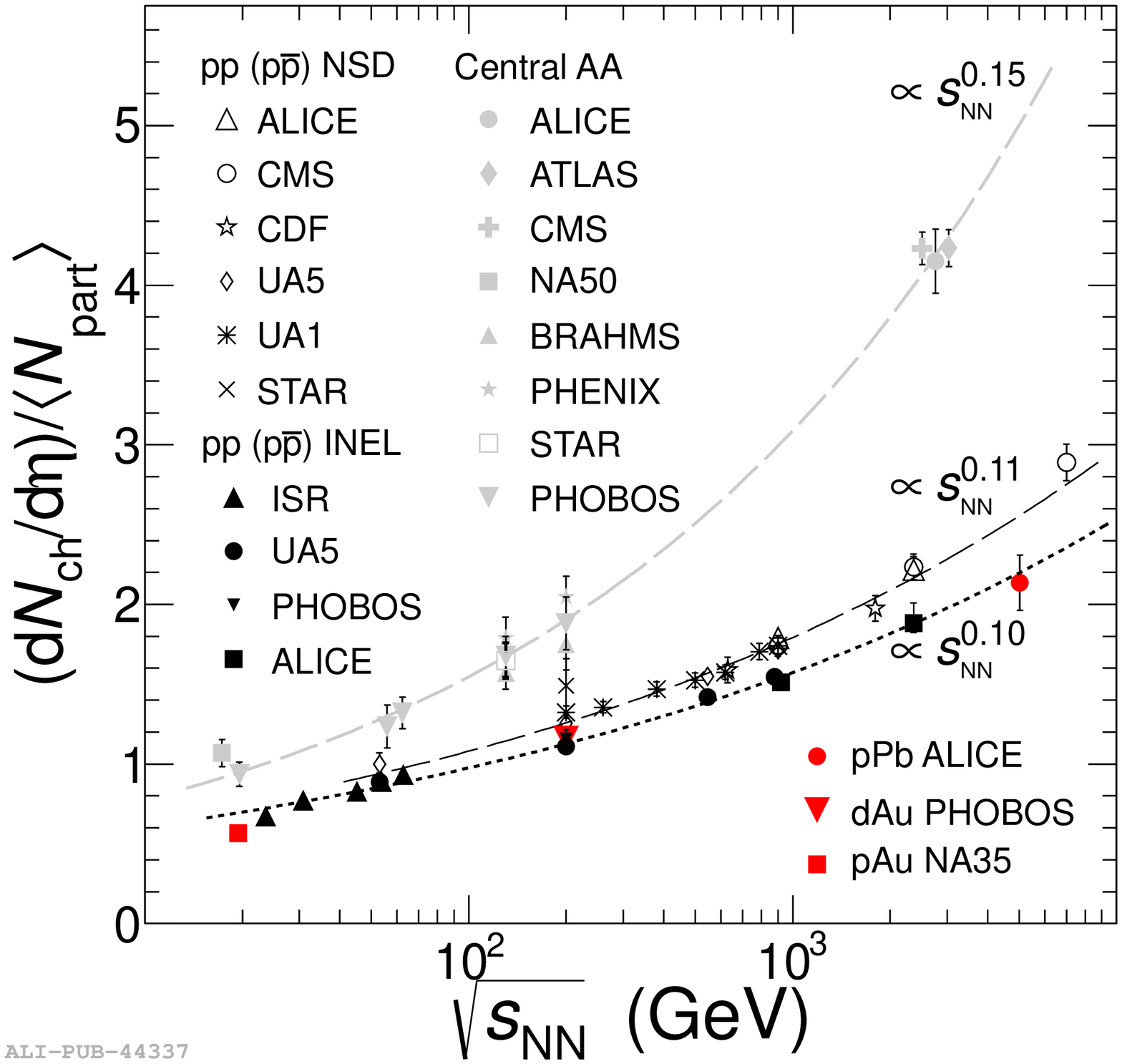}
	\end{subfigure}
\caption{Left: pseudorapidity density for Pb--Pb collisions at \sNN =~5.02 TeV in the centrality range 0--80\%, as a function of \avNpart~in each centrality class \cite{Adam:2015ptt}. Right: charged-particle pseudorapidity density at midrapidity normalized to the number of participants as a function of \sNN~\cite{ALICE:2012xs}.}\label{fig:2}	
\end{figure}

A study regarding the centrality determination in \mbox{p--Pb} collisions was carried out by ALICE, since the fluctuations in the multiplicity, which are present in this small system, bias the determination of the number of participants. A hybrid method is now used, based on the assumption of multiplicity at mid-rapidity being proportional to the number of participants \cite{Adam:2014qja}.
Results from Run 1 at 5.02 TeV collision energy are published, and Fig. \ref{fig:2} (right) shows the pseudorapidity density scaled by the number of participants as a function of the energy in the center of mass \cite{ALICE:2012xs}. The points are fitted with with a power law of the form $a\cdot s^{b} $ and one can notice that the pA points fit with the pp INEL class and not with the non-single diffractive sample. This is due to the negligibility of the diffraction contribution in a system like pA where more nucleons are involved. Another interesting fact comes from the observation of a much steeper rise of the AA points with respect to pp and pA. 

\section{Summary and outlook}
\label{summary}
Detailed studies for pp collisions at Run 1 and 2 have been presented, showing good predicting power extrapolating \dndetape~vs.~\sNN~from lower to higher energies~\cite{Adam:2015gka}.
There were no major surprises going from \sNN = 2.76 to 5.02 TeV in Pb--Pb collisions, distributions ratios scale flatly.
New multiplicity results are in preparation for p--Pb collisions at 8.16 TeV, and are expected to be published this autumn. They will benefit from a solid methodology elaborated in Run 1 to select the centrality of the event.

\end{document}